\documentclass{iaus}
\usepackage{graphicx}

\newcommand       \mum        {\,{\rm \mu m}}
\newcommand       \ohm        {\,{\rm ohm}}
\newcommand       \cm        {\,{\rm cm}}
\newcommand       \simali    {\,{\sim}}
\def    \Ftwenty     {F_{\rm 21\mu m}}
\def    \Fthirty     {F_{\rm 30\mu m}}
\def    \Fuir     {F_{\rm UIR}}
\def    \Fir      {F_{\rm IR}}

\title[Unidentified Species in Carbon Stars] 
{Unidentified species in envelopes around carbon stars}

\author[B.W.~Jiang et al.]   
{B.W.~Jiang$^1$,
  Aigen Li$^2$,
 Ke Zhang$^3$,
 J.M.~Liu$^1$,
 J.~Gao$^1$,
 \and A.~Mishra$^2$}

\affiliation{$^1$Department of Astronomy,
                 Beijing Normal University,
                 Beijing 100875, China\\
                 email:{\tt bjiang@bnu.edu.cn} \\[\affilskip]
             $^2$Department of Physics and Astronomy,
                 University of Missouri,
                 Columbia, MO 65211, USA \\[\affilskip]
             $^3$Department of Physics,
                 California Institute of Technology,
                 Pasadena, CA 91125, USA}

\pubyear{2013}
\volume{297}  
\pagerange{000--000}
\jname{Diffuse Interstellar Bands}
\editors{J.~Cami \& N. Cox, eds.}
\begin{document}

\maketitle

\begin{abstract}
%
The infrared (IR) spectra of many evolved carbon-rich stars
exhibit two prominent dust emission features
peaking around 21$\mum$ and 30$\mum$,
with the former exclusively seen in
proto-planetary nebulae (PPNe),
while the latter seen in a much wider range of objects,
including AGB stars, PPNe and planetary nebulae (PNe).
The 30$\mum$ feature is seen in all the 21$\mum$ sources,
but no correlation is found between these two features.
%
Over a dozen carrier candidates have been proposed
for the 21$\mum$ feature, but none of them has been widely
accepted and the nature of the 21$\mum$ feature remains
a mystery.
%
The carrier of the 30$\mum$ feature also remains unidentified.
MgS dust, once widely accepted as a valid carrier,
was ruled out because of the sulfur budget problem.
%
%
In this work we examine nano-sized FeO dust as a carrier
for the 21$\mum$ feature. We calculate the IR emission
spectrum of FeO nanodust which undergoes single-photon
heating in PPNe. It is found that the 21$\mum$ feature
emitted by FeO nanodust is too broad to explain
the observed feature.
For the 30$\mum$ feature, we argue that graphite could
be a viable carrier.
Graphite, provided its d.c. conductivity $\sigma_{\rm d.c.}$
exceeding $\simali100\ohm^{-1}\cm^{-1}$,
exhibits a pronounced band at 30$\mum$.

\keywords{stars: carbon, (stars:) circumstellar matter,
          dust, infrared: stars}
\end{abstract}

\firstsection 
\section{Introduction}
In the late stage of evolution, the low- and intermediate-mass stars
lose large amount of mass and produce circumstellar shells.
Various dust species are present in the cold and dense shells
of asymptotic giant branch (AGB) stars,
post-AGB stars and planetary nebulae (PNe),
among which silicate dust in O-rich shells
and SiC dust in C-rich shells
are well identified through their spectral features.

There are two prominent spectral features
respectively peaking around 21$\mum$ and 30$\mum$.
Both features are observed in circumstellar shells
of C-rich stars in both the Milky Way
and the Magellanic Clouds (MC).
So far, the 21$\mum$ feature is detected
in about 18 Galactic objects (Cerrigone et al.\ 2011)
and 9 MC objects (Volk et al.\ 2011).
The 21$\mum$ feature is only seen
in proto-planetary nebulae (PPNe),
a transient phase in stellar evolution.
In contrast, the 30$\mum$ feature is detected
in the whole late stage of stellar evolution,
from AGB to post-AGB to PN (Hony et al.\ 2003).

\begin{figure}[h]
 \vspace*{-1.0cm}
\begin{center}
 \hspace*{-0.8cm}
 \includegraphics[width=6.0in]{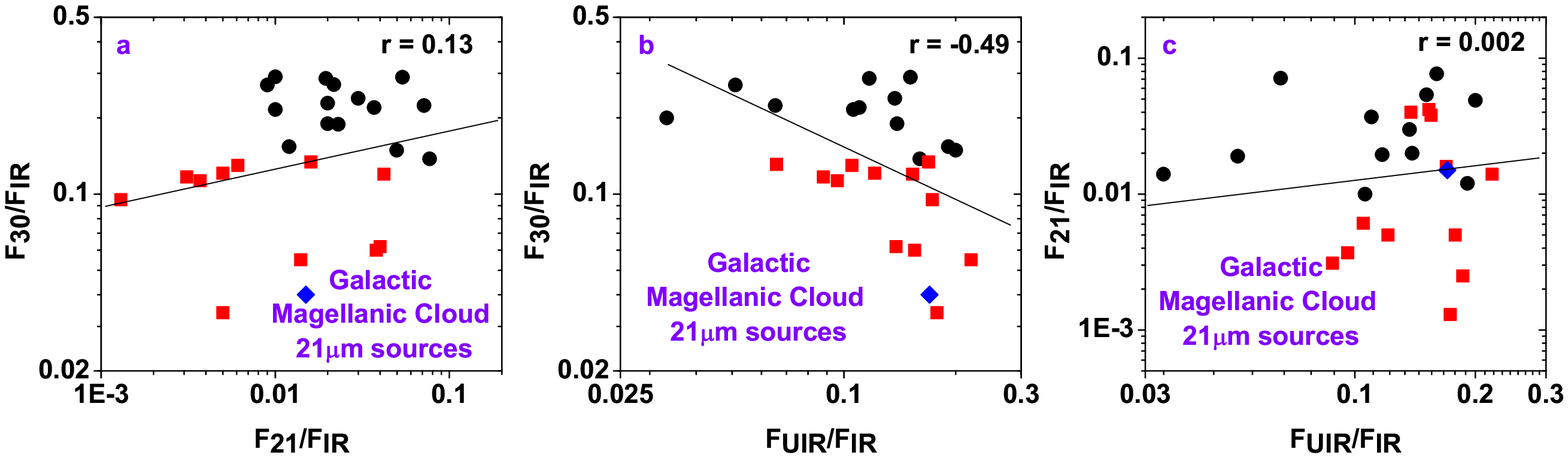}
 \vspace*{-1.0cm}
\caption{\footnotesize
         \label{fig:21vs30}
         Interrelations among the 21$\mum$, 30$\mum$,
         and UIR features of Galactic (circles)
         and MC (squares) 21$\mum$ sources:
         (a) $\Fthirty$ vs. $\Ftwenty$,
         (b) $\Fthirty$ vs. $\Fuir$, and
         (c) $\Ftwenty$ vs. $\Fuir$.
         All quantities are normalized by $\Fir$,
         the total near- to mid-IR emission obtained
         by {\it ISO}/SWS in the 2--45$\mum$ wavelength range,
         to cancel out their common proportionality to $\Fir$
         (i.e. the illuminating starlight intensity
         and the bulk dust quantity).
         }
\end{center}
\vspace*{-0.3cm}
\end{figure}

All the 21$\mum$ sources also display
the 30$\mum$ emission feature
as well as the so-called ``unidentified infrared (UIR)''
emission bands at 3.3, 6.2, 7.7, 8.6 and 11.3$\mum$.
The UIR bands are commonly attributed to polycyclic
aromatic hydrocarbon (PAH) molecules
(L\'eger \& Puget 1984, Allamandola et al.\ 1985).
As shown in Figure~\ref{fig:21vs30},
the 30$\mu$m feature does not correlate with
the 21$\mu$m feature, implying that their carriers
are not related.
Figure~\ref{fig:21vs30} also shows that the 21$\mum$ feature
does not correlate with the UIR features.
This argues against large PAH clusters
as a possible carrier for the 21$\mum$ feature.
Moreover, it appears that the 30$\mum$ feature
and the UIR features weakly anti-correlate,
suggesting that the UIR carriers (e.g. PAHs)
may result from the decomposition or shattering of
the 30$\mum$-feature carrier.

\begin{figure}[h]
\begin{center}
 \includegraphics[width=3.4in]{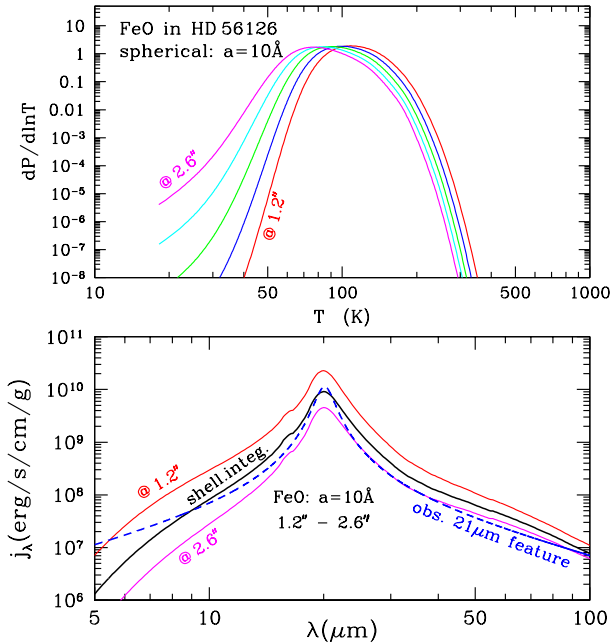}
\caption{\footnotesize
         \label{fig:FeO}
        Upper panel: the temperature probability distribution
        functions $dP/dT$ of FeO dust of $a$\,=\,1\,nm
        at various distances from the illuminating star
        HD\,56126.
        Lower panel: The emission spectra of FeO dust
        of $a$\,=\,1\,nm at the inner and outer shells
        (i.e., 1.2$^{\prime\prime}$ and 2.6$^{\prime\prime}$
         from the central star), as well as that obtained
         from integrating the whole dust shell.
         Also shown is the observed 21$\mum$ emission feature.
         }
\end{center}
\vspace{-5mm}
\end{figure}

\begin{figure}[h]
\begin{center}
 \includegraphics[width=3.4in]{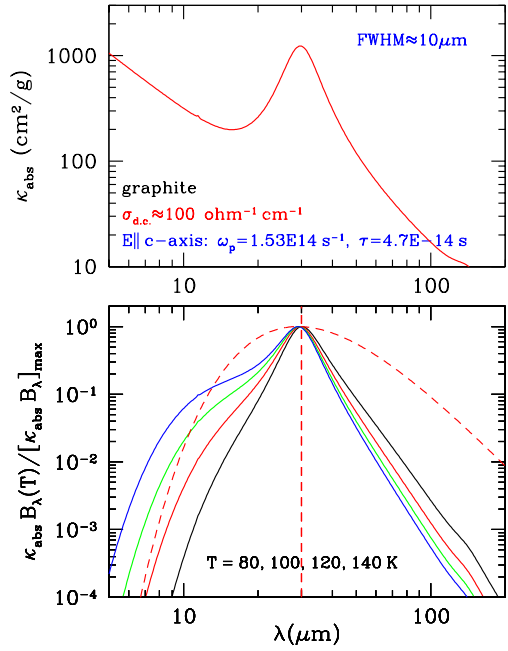}
\caption{\footnotesize
         \label{fig:30um}
         Upper panel: the opacity of graphite
         with a d.c. conductivity of
         $\sigma_{\rm d.c.} = 100\ohm^{-1}\cm^{-1}$.
         A prominent 30$\mum$ feature with
         a FWHM of $\simali$10$\mum$ is
         clearly seen.
         Lower panel: the emission profiles
         of graphite with
         $\sigma_{\rm d.c.}=100\ohm^{-1}\cm^{-1}$
         at  $T$\,=\,80, 100, 120, 140\,K.
         Dot-dashed line: the $T$\,=\,100\,K blackbody.
         }
\end{center}
\vspace*{-0.1cm}
\end{figure}

\section{FeO nanodust as a carrier of the 21$\mum$ feature}
Zhang et al.\ (2009a) examined nine inorganic carrier candidates
for the 21$\mum$ feature, including nano TiC, fullerenes with Ti,
SiS$_{2}$, doped SiC, the SiC and SiO$_{2}$ mixture,
FeO, Fe$_{2}$O$_{3}$ and Fe$_{3}$O$_{4}$.
We found that except FeO nanodust, they are all problematic:
they either require too much material
(i.e., violating the abundance constraint)
or produce extra features which are not seen
in astronomical objects.

FeO seems to be a promising candidate carrier for
the 21$\mum$ feature for two reasons:
(1) Fe is an abundant element; and
(2) FeO has a pronounced spectral feature around 21$\mum$
(Posch et al.\ 2004, Zhang et al.\ 2009a).
However, it is not clear how FeO forms in
C-rich environments as the 21$\mum$ sources
are all C-rich. Posch et al.\ (2004) argued that FeO nanodust
could form in these environments.
We therefore model the excitation and emission
processes of FeO nanodust in PPNe. As FeO nanodust
will experience temperature fluctuation, we calculate
its temperature probability distribution function $dP/dT$
and then calculate its emission spectrum from $dP/dT$.

Taking a 3-D Debye model (with a Debye temperature of
$\Theta_{\rm D}$\,=\,430\,K) for the specific heat of FeO
and the refractive indices at T\,=\,10\,K, 100\,K, 200\,K
and 300\,K from Henning \& Mutschke (1995, 1997),
we calculate $dP/dT$ and the emergent spectrum of
spherical FeO of radius $a$\,=\,10\,{\AA}
for HD\,56126, a proto-typical 21$\mum$ source
(see Figure~\ref{fig:FeO}).
The IR emission spectrum exhibits a strong band at
21$\mum$. However, it is too broad
(with FWHM\,$\simali$4$\mum$) to explain
the observed FWHM of $\simali$2.2--2.3$\mum$
(see Kwok et al.\ 1989).
For nonspherical dust, the predicted 21$\mum$
feature is even broader.
Finally, to account for the power emitted
from the 21$\mum$ feature in HD\,56126,
the FeO model requires
${\rm Fe/H} \approx 4.3\times10^{-6}$,
exceeding the iron budget of
${\rm Fe/H} \approx 3.2\times10^{-6}$
available in HD\,56126.
Therefore, FeO nanodust is unlikely
a valid carrier for the 21$\mum$ feature.

\section{Graphite as a carrier for the 30$\mum$ feature}
The 30$\mum$ feature,
extending from $\simali$24 to $\simali$45$\mum$,
is very strong and accounts for up to $\simali$30\%
of the total IR luminosity of the emitter.
MgS dust, once widely accepted as a carrier
(Begemann et al.\ 1994), was ruled out
by Zhang et al.\ (2009b) who showed that the MgS model
would require too much S.
Even with a very generous assumption of UV/optical
absorbing coefficient, the required MgS dust
exceeds what is available by over an order of magnitude.
This was later confirmed by
Messenger et al.\ (2013)
(but also see Lombaert et al.\ 2012).

We propose graphite as a carrier of the 30$\mum$ feature
seen in evolved stars. Graphite has been identified
as a presolar stardust species
in primitive meteorites through its
isotropic anomaly. Graphite is also key component
in standard interstellar grain models.
A feature around 30$\mum$ is already recognizable in
the absorption coefficient of graphite (Draine \& Lee 1984).
Depending on its d.c. conductivity $\sigma_{\rm d.c.}$,
the 30$\mum$ feature could be very strong.
Our preliminary calculations show that graphite could
account for the observed 30$\mum$ feature,
provided $\sigma_{\rm d.c.} > 100\ohm^{-1}\cm^{-1}$
(see Figure~\ref{fig:30um}).
Experiments have shown that graphite can have a wide range
of $\sigma_{\rm d.c.}$,
ranging from $\simali$1
to $\simali$200$\ohm^{-1}\cm^{-1}$ (Primak 1956).

\acknowledgments{%
We are supported in part by
NSF AST-1109039, NNX13AE63G,
NSFC\,11173007, NSFC\,11173019,
and the University of Missouri Research Board.
}

\end{document}